\documentclass{eas}
\usepackage{graphicx}
\usepackage{amssymb}

\begin{document}

\title{Is Palomar 1 really associated with the Canis Major overdensity? }
\runningtitle{Palomar 1 and the CMa overdensity}
\author{Ivo Saviane}
\address{European Southern Observatory, Chile}
\author{Lorenzo Monaco}
\sameaddress{1}
\author{Matteo Correnti}
\address{Istituto Nazionale
di Astrofisica \textendash{} Osservatorio Astronomico di Bologna,
Italy}
\author{Piercarlo Bonifacio}
\address{GEPI, Observatoire de Paris, CNRS, Universite Paris Diderot,
France}
\secondaddress{Istituto Nazionale di Astrofisica \textendash{} Osservatorio
Astronomico di Trieste, Italy}
\author{Doug Geisler}
\address{Universidad de Concepcion, Chile} 

\begin{abstract}
It has recently been suggested that the peculiar cluster
Pal~1 is associated to the Canis Major dwarf galaxy, which existence
is still at the center of a debate. Our first measurement of the 
cluster's  chemical abundance ratios allows us to examine this 
association and to advance further in the clarification of Pal~1
possible origin.
\end{abstract}
\maketitle
\section{Introduction}
\begin{figure}
\noindent \begin{centering}
\includegraphics[width=0.99\columnwidth]{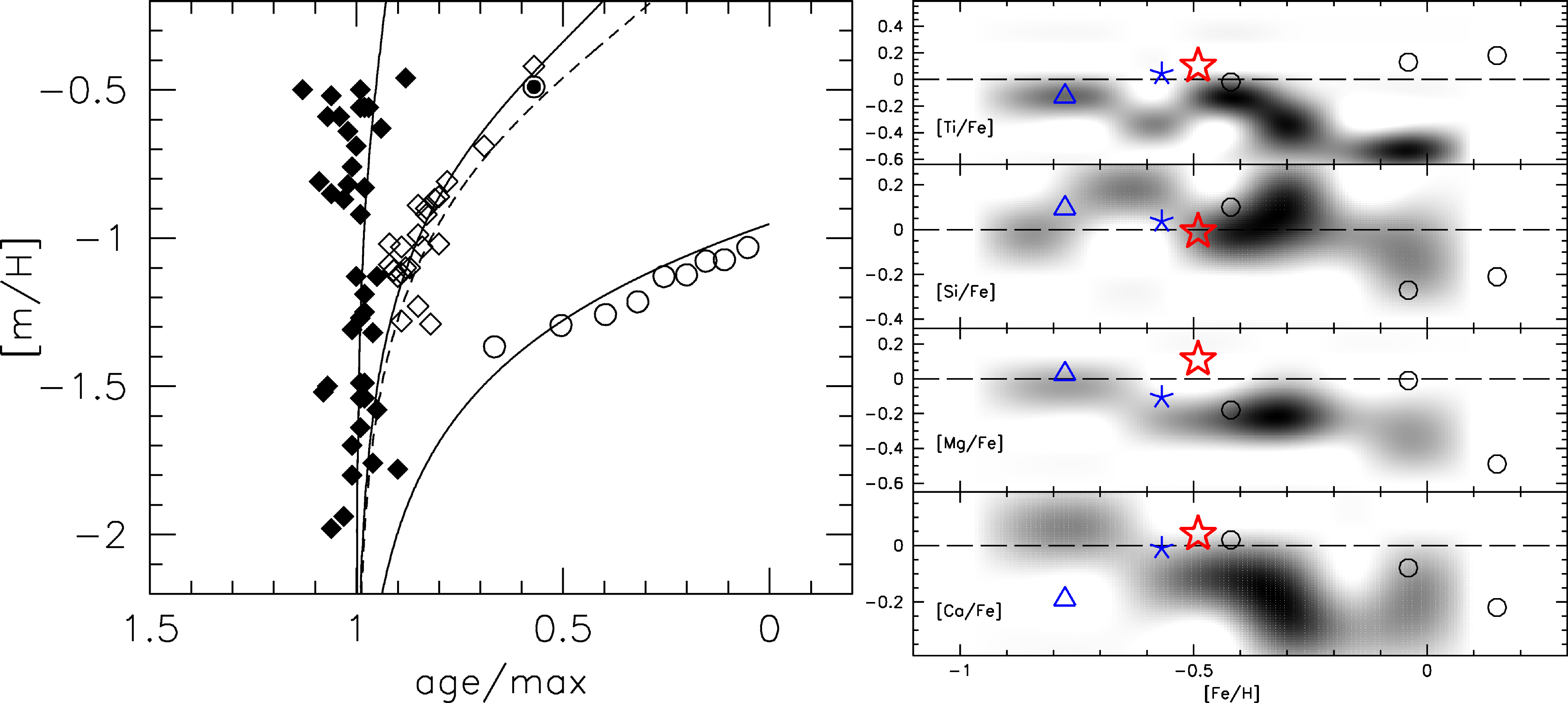}
\par\end{centering}

\caption{{\bf Left panel} -- The filled and open diamonds represent the
  AMR of old- and young-halo globular clusters from Marin-Franch et al.
  (\cite{marin-franch}), while open circles represent the Leo~I AMR from
  Gullieuszik et al.  (\cite{marco09}).  The curves are three closed-box
  models with a yield $y=Z_{\odot}/2$ and different gas consumption
  rates $d\mu/dt$, where $\mu=M_{{\rm gas}}/M_{{\rm tot}}$: $d\mu/dt=10$
  for old clusters, $d\mu/dt=1.2$ for young clusters, and
  $d\mu/dt=0.2$ for Leo~I. The dashed curve corresponds to
  the AMR of the Sgr dSph (Siegel et al.  \cite{siegel-etal07}). The
  encircled dot represents Pal~1.  {\bf Right panel} -- Pal1-I (big
  empty red star) $\alpha$-element abundance ratios are presented
  together with Sagittarius dSph stars (grey shaded area) from Sbordone
  et al. (\cite{sbordone}), and stars in the GCs Pal~12 (big open blue
  triangle) and Ter~7 (big blue asterisk) from Cohen (\cite{cohen04})
  and Sbordone et al. (\cite{sbordone}), respectively.  Stars in the CMa
  overdensity region in the background of the open cluster NGC\,2477 are
  plotted as big open circles (Sbordone et al. \cite{sbordone05}).
  \label{fig:Here-we-show-amr}}

\end{figure}

With an age of $\sim8$~Gyr and a metallicity {[}Fe/H{]}$\simeq-0.6$,
Pal~1 does not fit the age-metallicity relation (AMR) of classical
old-halo globular clusters (Rosenberg et al. \cite{R98a,R98b}). It is
compatible with the AMR of younger globulars, which is also similar to
that of the Sgr dwarf spheroidal (dSph) galaxy (Siegel et al.
\cite{siegel-etal07}).  The putative Canis~Major dwarf was also
suggested to have an AMR similar to that of Sgr (Forbes \& Bridges
\cite{forbes_bridges}), and Pal\,1 lies close to the CMa orbit in phase
space (Martin et al.  \cite{martin}). %
The AMRs shown in Fig.~\ref{fig:Here-we-show-amr} (left panel) define a
trend, with higher mass galaxies having larger gas consumption rates
$d\mu/dt$.  Leo~I ($\sim2\times10^{6}M_{\odot}$, Mateo \cite{m98}) has a
rate smaller than that of Sgr ($\gtrsim 10^{9}M_{\odot}$, Jiang \&
Binney \cite{jiang}), whose rate is, in turn, smaller than that of the
(proto) Galaxy.  The graph therefore suggests that Pal~1 was created in
a galaxy with a mass similar to that of Sgr.  If Sgr and CMa share a
similar AMR, then Fig.~\ref{fig:Here-we-show-amr} would suggest that
they also have similar masses and star formation (SF) efficiencies, and
presumably also similar abundance ratios, which should be reflected in
the Pal~1 ratios, most notably in the $\alpha$-elements.  To check this
possibility, these ratios were measured using a high-resolution spectrum
of star Pal~1-I (see Rosenberg et al.  \cite{R98b}) obtained with
HDS@SUBARU (Monaco et al., 2010, A\&A, in press; arXiv1011.0123).

\section{Abundance ratios of Pal 1}

The $\alpha$-element abundance ratios of Pal~1 are presented in
Fig.~\ref{fig:Here-we-show-amr} (right panel) together with those of Sgr stars and its
associated clusters Pal~12 and Ter~7 (see Cohen \cite{cohen04} and 
Sbordone et al. \cite{sbordone}), and of stars in the CMa overdensity region
(Sbordone et al. \cite{sbordone05}).
Abundances of all other analyzed elements are presented in Monaco et al.
The most metal-poor among CMa stars has an iron content similar to that
of Pal\,1-I and the chemical composition of the two objects is also
similar.
In general, Sbordone et al. (\cite{sbordone05}) suggested that the CMa
structure has a level of chemical processing compatible with that of the
Galactic disk, true also for Pal~1. An association of the cluster to CMa
might then be possible\footnote{It should be noted, however, that the
  low Galactic latitude of CMa makes contamination from the disk itself
  a relevant issue, and make it difficult to unambiguously assign
  membership of the stars studied by Sbordone et al.
  (\cite{sbordone05}).  }.  The abundance ratios of Sgr are instead
quite different. While the abundance of silicon is compatible to that of
Sgr, the rest of $\alpha$-elements (Fig.~\ref{fig:Here-we-show-amr}),
iron-peak and light element abundance ratios (see Monaco et al. 2010)
are distinctively different.

In conclusion, Pal~1 chemical abundance pattern is similar to that of
CMa (albeit based on one star only) but also to Galactic open clusters (OCs). The
cluster might therefore be associated to CMa, although the separation
between CMa and the Galactic disk remains an open question.  

Contrary to the expectations of the Introduction, CMa and Sgr have
different abundance patterns despite having the same AMR.  Sgr does have
an $\alpha$-element pattern similar to those of other dSph of the Local
Group (Monaco et al. 2005; Sbordone et al. 2007), so one solution of
this puzzle might be that CMa is not really an independent entity, and
the mass-SF~efficiency relation does not apply to it (Momany et al.
\cite{yaz}).

Palomar~1 might then be a GC which experienced a peculiar chemical
evolution or an OC ejected from the Galactic disk, which high
concentration and flat mass function (see Rosenberg et al. \cite{R98b})
would be the result of its peculiar dynamical evolution.  Only a
reconstruction of the cluster's orbit would permit discriminating
between the two possibilities. Because radial velocities and positions
are not sufficient to claim a common orbit, a measurement of proper
motions is needed to break degeneracies.


\begin{thebibliography}{99}

\bibitem[2004]{cohen04} Cohen, J. G. 2004, AJ, {127}, 1545

\bibitem[2010]{forbes_bridges} Forbes, D. A., \& Bridges, T. 2010, MNRAS,
  {404}, 1203

\bibitem[2009]{marco09} Gullieuszik, M., Held, E. V., Saviane, I., \& Rizzi,
  L. 2009, A\&A, {500}, 735


\bibitem[2000]{jiang} Jiang, I.-G., \& Binney, J. 2000, MNRAS, {314}, 468

\bibitem[2009]{marin-franch} Marin-Franch, A., et al. 2009, ApJ, {694}, 1498

\bibitem[2004]{martin} Martin, N. F., et al 2004, MNRAS, {348}, 12

\bibitem[1998]{m98}Mateo, M. L. 1998, ARA\&A, 36, 435

\bibitem[2006]{yaz} Momany, Y., et al.  2006. A\&A, 451, 515

\bibitem[2005]{monaco_etal05} Monaco, L., et al.  2005, A\&A, 441, 141


\bibitem[1998a]{R98a} Rosenberg, A., et al. 1998a, AJ, {115}, 648

\bibitem[1998b]{R98b} Rosenberg, A., Piotto, G., Saviane, I., Aparicio, A., \&
  Gratton, R. 1998b, AJ, {115}, 658

\bibitem[2005]{sbordone05} Sbordone, L., et al. 2005, A\&A, {430}, L13


\bibitem[2007]{sbordone} Sbordone, L., et al. 2007, A\&A, {465}, 815



\bibitem[2007]{siegel-etal07} Siegel, M. H., et al 2007., ApJ, {667}, L57


\end{thebibliography}
\end{document}